\documentclass[%
 reprint,
superscriptaddress,
 amsmath,amssymb,
 aps,
 prl
]{revtex4-1}

\usepackage{graphicx}
\usepackage{dcolumn}
\usepackage{bm}
\usepackage{threeparttable}
\usepackage{url} 
\usepackage[colorlinks,linkcolor=blue,anchorcolor=blue,citecolor=blue,urlcolor=blue]{hyperref}

\usepackage[mathlines]{lineno}
\begin{document}


\title{Unusual Transformation of Polymer Coils in a Mixed Solvent Close to the Critical Point}

\author{Xiong Zheng}
\affiliation{Institute for Physical Science and Technology and Department of Chemical and Biomolecular Engineering, University of Maryland, College Park, Maryland 20742, United States}
\affiliation{Key Laboratory of Thermo-Fluid Science and Engineering, Ministry of Education, Xi'an Jiaotong University, Xi'an, Shaanxi Province 710049, P.R. China}

\author{Mikhail A. Anisimov}
\email{anisimov@umd.edu}
\affiliation{Institute for Physical Science and Technology and Department of Chemical and Biomolecular Engineering, University of Maryland, College Park, Maryland 20742, United States}

\author{Jan V. Sengers}
\affiliation{Institute for Physical Science and Technology and Department of Chemical and Biomolecular Engineering, University of Maryland, College Park, Maryland 20742, United States}

\author{Maogang He}
\affiliation{Key Laboratory of Thermo-Fluid Science and Engineering, Ministry of Education, Xi'an Jiaotong University, Xi'an, Shaanxi Province 710049, P.R. China}

\date{\today}

\begin{abstract}
We have discovered unusual behavior of polymer coils in a binary solvent (nitroethane+isooctane) near the critical temperature of demixing. The exceptionally close refractive indices of the solvent components make the critical opalescence relatively weak, thus enabling us to simultaneously observe the Brownian motion of the polymer coils and the diverging correlation length of the critical fluctuations. The polymer coils exhibit a collapse-reswelling-expansion-reshrinking transition upon approaching the critical temperature. While the first stage (collapse-reswelling) can be explained by the theory of Brochard and de Gennes, the subsequent expansion-reshrinking transition is a new unexpected phenomenon that has not been observed so far. We believe that this effect is generic and attribute it to micro-phase separation of the solvent inside the polymer coil. 
\begin{description}
\item[PACS numbers]
05.70.JK, 61.25.he
\pacs{Valid PACS appear here}
\keywords{Suggested keywords}
\end{description}
\end{abstract}

\maketitle

     Collapse of polymer coils in a poor solvent and its swelling in a good solvent is undoubtedly one of the most famous phenomena in polymer science \cite{flory1953principles,williams1981polymer}. In addition, Brochard and de Gennes have predicted collapse and reswelling (to the original size) of a polymer coil in a mixture of good solvents near the consolute critical point due to the growing correlation length of the critical fluctuations and preferential adsorption of the better solvent \cite{williams1981polymer,brochard1980collapse}. This effect has been qualitatively confirmed by computer simulations \cite{magda1988dimensions, luna1997polymer, vasilevskaya1998conformation, dua1999polymer, sumi2005cooperative, sumi2007behavior,sumi2009critical} and by three experiments \cite{to1998polymer, grabowski2007contraction, he2012partial}. However, there is still an important question regarding the adequacy of the theory of Brochard and de Gennes. An essential feature of the theory of Brochard and de Gennes is the assumption that the polymer is well soluble in both mixture components. In this case the critical temperature would decrease with increasing polymer concentration for an upper critical solution point and would increase for the case of a lower critical solution point, thus always shrinking the two-phase domain. Since demixing of the solvent means a significant difference in the molecular interactions of the components, one component is usually a significantly better solvent for a polymer than the other one. Therefore, this assumption is not commonly valid. Indeed, in the systems investigated in Refs. \cite{to1998polymer, grabowski2007contraction, he2012partial} the two-phase liquid-liquid domain increases with the addition of a polymer so that they do not satisfy the assumption of Brochard and de Gennes. Furthermore, in a previous fluorescence-correlation spectroscopy study of the hydrodynamic radius of a polymer coil \cite{grabowski2007contraction} and in a small-angle neutron scattering study of the radius of gyration \cite{he2012partial}, the opalescence was experimentally eliminated, so that the correlation length of the critical fluctuations in the polymer solution could not be compared with the dimensions of polymer coils.\par
     The initial motivation of our study was to quantitatively verify the theory of Brochard and de Gennes for a typical near-critical binary liquid, in which one component would be a good solvent or a theta solvent, while the second component would be a poor solvent. Moreover, we intended to establish a relation between the correlation length and the coil dimensions. However, we have also discovered a new unexpected phenomenon that has not been observed so far.\par
     In our work the exceptionally close difference between the refractive indices of the solvent components, nitroethane and isooctane (\(\sim\)0.002), makes the critical opalescence relatively weak (while still reliably measurable very close to the critical point) \cite{aref1973mandel}, thus enabling us to simultaneously observe the Brownian motion of the polymer coils and the diverging correlation length of the critical fluctuations.\par
     Four polymer samples were used in the study: polystyrene 25,000 $\textup{g mol}^{-1}$ with $M_{\textup{w}}$/$M_{\textup{n}}$=1.06 (PS-25); polystyrene 50,000 $\textup{g mol}^{-1}$ with $M_{\textup{w}}$/$M_{\textup{n}}$=1.06 (PS-50); polystyrene 123,000 $\textup{g mol}^{-1}$ with $M_{\textup{w}}$/$M_{\textup{n}}$=1.08 (PS-123);  and poly(butyl methacrylate) 180,000 $\textup{g mol}^{-1}$ with $M_{\textup{w}}$/$M_{\textup{n}}$=1.12 (PBMA-180). Polystyrene was purchased from Alfa-Aesar and PBMA from Scientific Polymer. Analytical grade nitroethane and isooctane with a purity of 99.5\% were purchased from Alfa-Aesar. All samples were studied at the critical concentration, 53.5\% mass of isooctane \cite{beysens1979coexistence}. The solutions were filtered through Millipore 20 nm filters at about 5 $^{\circ}$C above the critical temperature.\par
     A dynamic light-scattering technique was used to measure both the hydrodynamic radius of polymer coils in the dilute solutions (away from and close to the critical point) and the correlation length (close to the critical point). The method and the instruments have been described elsewhere \cite{jacob2001light, anisimov2005competition}. All measurements were performed at a scattering angle of 20$^{\circ}$. The temperature was stabilized within about \(\pm\)0.003 $^{\circ}$C.\par
     In the close vicinity of the consolute critical temperature in the presence of polymers, double-exponential time-dependent autocorrelation functions, \(g(\tau)\), were found:
\begin{equation}
g(\tau)-1=[A_1\textup{exp}(-\Gamma_1\tau)+A_2\textup{exp}(-\Gamma_2\tau)]^2,
\end{equation}
where \(\Gamma_1\) and \(\Gamma_2\) are the decay rates. In the mixture without polymer, only single-exponential correlation functions were observed. The hydrodynamic radius of the polymer chains, \(R_\textup{h}\) was calculated from the Stokes-Einstein relation
\begin{equation}
R_\textup{h}=\frac{k_\textup{B}T}{6\pi\eta q^2}\Gamma_1,
\end{equation}
while the correlation length of the critical fluctuations was calculated from the mode-coupling  equation (see Ref. \cite{jacob2001light}):
\begin{equation}
\xi=\frac{k_\textup{B}T}{6\pi\eta q^2}\Gamma_2K(q\xi).
\end{equation}\par
In Eqs. (2) and (3) \(\textit{k}_\textup{B}\), \textit{T}, \(\eta\) and \(\textit{q}\) are the Boltzmann constant, temperature, viscosity of  the solvent, and the wavenumber, respectively, \(\textit{q}=(4\pi\textit{n}/\lambda)\textup{sin}(\theta/2)\), \(\lambda\) is the wavelength of the incident light in vacuum, \textit{n} is the refractive index, \(\theta\) is the scattering angle. The function \(\textit{K}(\textit{q}\xi\)), is the universal “Kawasaki function” \(\textit{K}(\textit{q}\xi)=3/4\{1+(\textit{q}\xi)^{-2}+[\textit{q}\xi-(\textit{q}\xi)^{-3}]\}[\textup{tan}^{-1}(\textit{q}\xi)]\) \cite{jacob2001light}. The viscosity and refractive indices of the mixed solvent were taken from Refs. \cite{beysens1979coexistence,garrabos1982high}, the refractive indices of nitroethane and isooctane are from Refs. \cite{toops1956physical,zhang2015densities}. A more detailed analysis of experimental data is given in Section 1 of the Supplemental Material \cite{SupplementalMaterial}.\par
\begin{figure*} 
\centering
\includegraphics[width=5.4in,height=4.3in]
{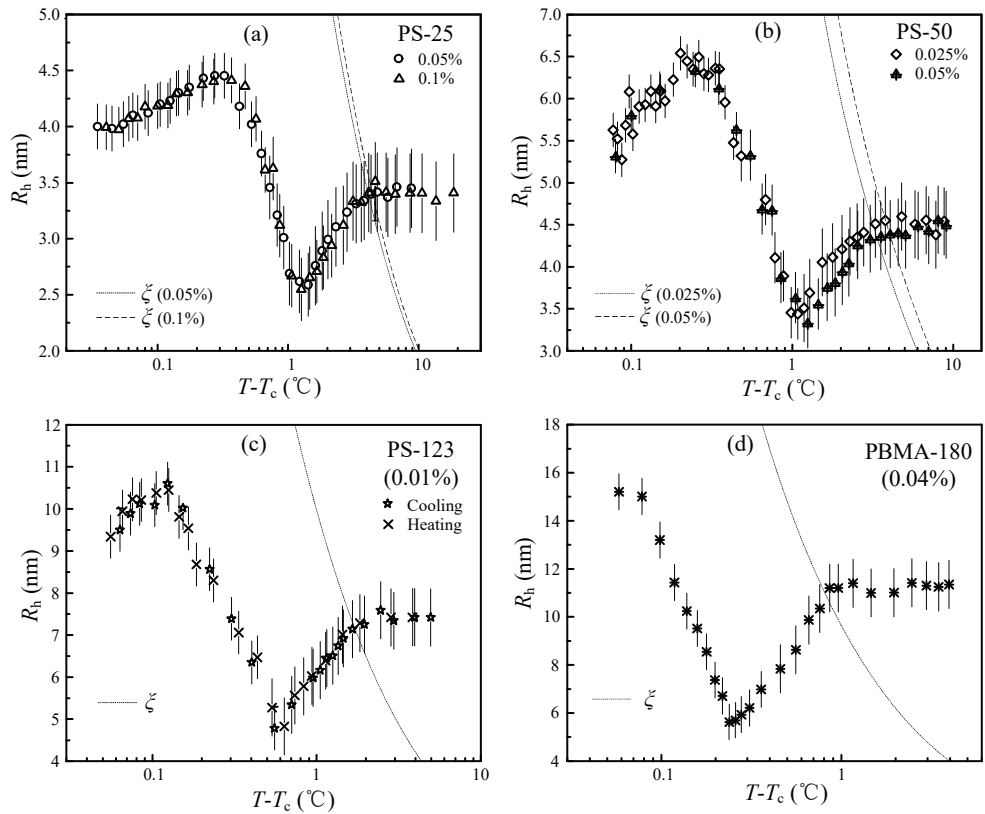}
\caption{\label{fig:Fig1}The dependence of the hydrodynamic radius \(R_\textup{h}\) of polymer coils on the temperature distance from the critical point. (a) PS-25 (0.05\% and 0.01\% mass); (b) PS-50 (0.025\% and 0.05\% mass); (c) PS-123 (0.01\% mass); (d) PBMA-180 (0.04\% mass). The behavior of the correlation length \(\xi\) is shown by the corresponding dotted and dashed curves.}
\end{figure*}

     The hydrodynamic radii of polymer chains as a function of the temperature distance from the critical point are shown in Fig. \ref{fig:Fig1}. For all polymer solutions studied, the coil size does not change, when the correlation length is smaller than the hydrodynamic radius of a polymer coil. When the correlation length becomes comparable with or larger than the hydrodynamic radius, the polymer chains collapse. Upon further increase of the correlation length, the polymer coils reswell. This collapse-reswelling behavior is consistent with the theory of Brochard and de Gennes \cite{brochard1980collapse}. However, unlike the prediction of this theory, the reswelling does not terminate at the original size. Instead, the coils expand beyond the original size and shrink again in the immediate vicinity of the critical point. The temperature of this unexpected expansion-reshrinking transition is shifted closer to the critical temperature with increase of the polymer molecular weight. This is why for PBMA-180 (Fig. \ref{fig:Fig1}(d)) and in earlier studies \cite{grabowski2007contraction,he2012partial} the final reshrinking to the original size was not fully observed. This new phenomenon is fully reproducible. In particular, independent runs upon cooling and heating (Fig. \ref{fig:Fig1}(c)) yield identical results within the uncertainty of the measurements.\par
\begin{figure} 
\centering
\includegraphics[width=2.8in,height=4.2in]{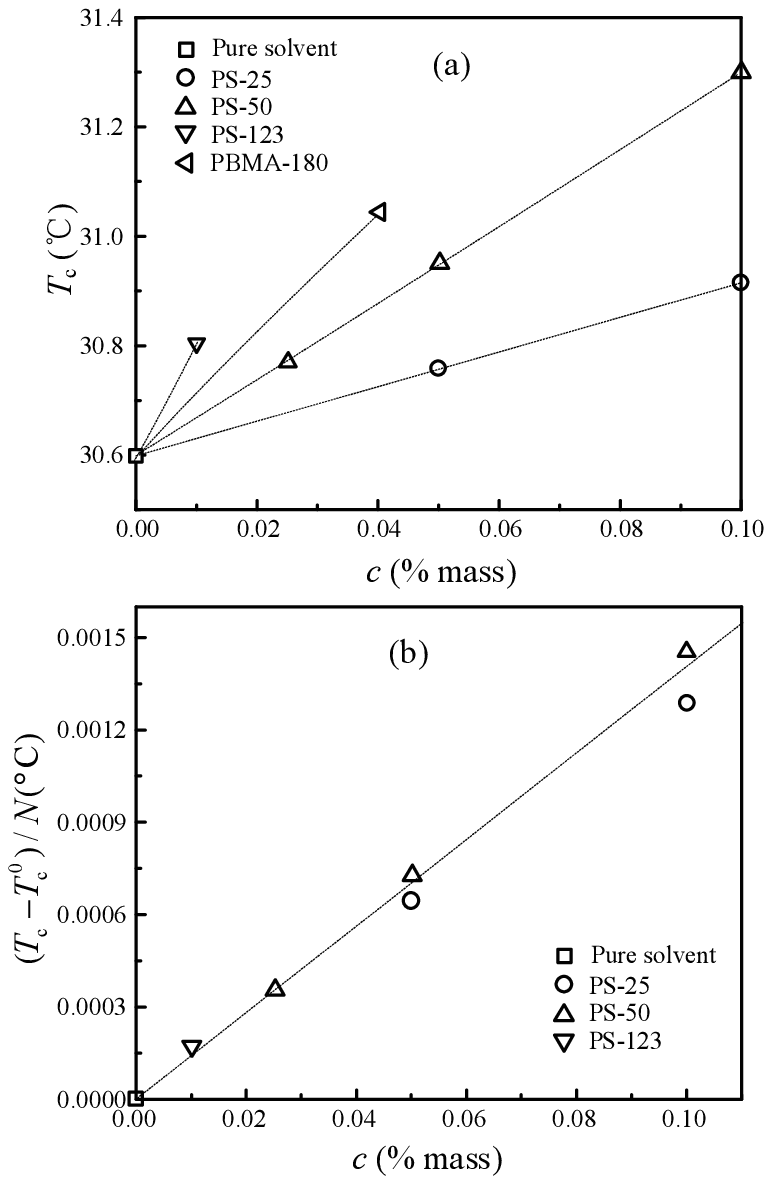}
\caption{\label{fig:Fig2}(a) Dependence of the critical temperature on polymer concentration; (b) universal scaled relationship between the shift of the demixing critical temperature in the PS solutions with respect to the value in the pure binary solvent reduced by the degree of polymerization (\textit{N}) and polymer concentration.}
\end{figure}

     The nitroethane+isooctane mixture has an upper critical solution temperature \cite{aref1973mandel,beysens1979coexistence} and the addition of a polymer increases the critical temperature, thus again contradicting the major assumption in the theory of Brochard and de Gennes. The behavior of the critical temperatures for the studied systems is shown in Fig. \ref{fig:Fig2}.\par
     We note that the difference between the critical temperature \(T_\textup{c}(c)\) of the polystyrene solutions and the critical temperature \(T_\textup{c}^0\) of the pure binary solvent, becomes a universal function of the polymer concentration \textit{c}, when this temperature difference is reduced by the degree of polymerization \textit{N} (the number of monomers in the coil). This universality is caused by the proportionality between the number of monomers in the polymer coil and the polymer-solvent interaction energy.\par
\begin{figure} 
\centering
\includegraphics[width=2.6in,height=4.2in]{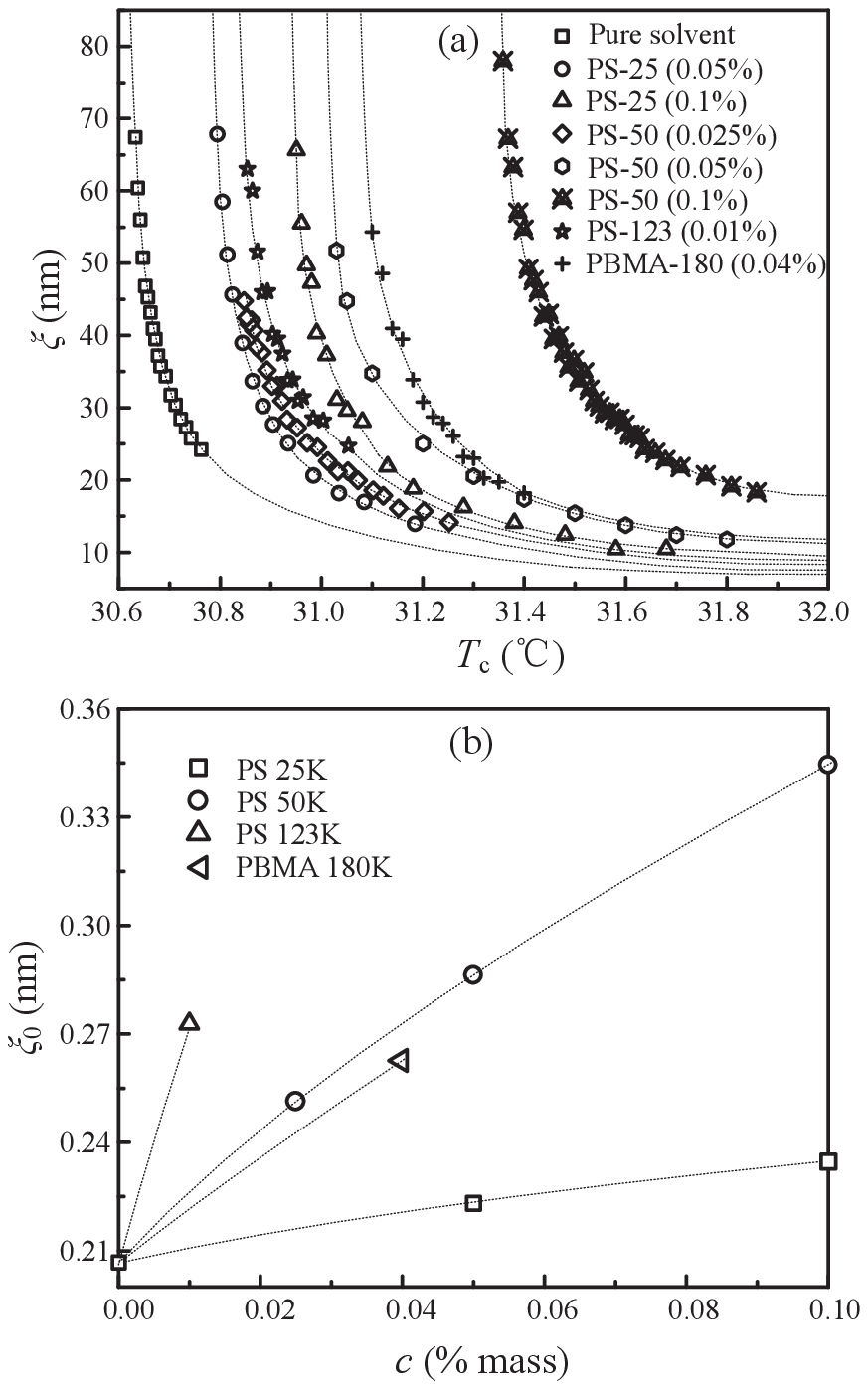}
\caption{\label{fig:Fig3}(a) Asymptotic divergence (\(\xi=\xi_0t^{-0.63}\), shown by dotted curves) of the critical correlation length near the critical temperatures of different polymer solutions; (b) amplitudes of the correlation length for the same solutions.}
\end{figure}

     The diverging correlation length of the critical fluctuations as a function of temperature for all the studied samples is shown in Fig. \ref{fig:Fig3}(a). We confirm the result of Garrabos \textit{et al}. \cite{garrabos1982high}, obtained from a static light-scattering experiment for the nitroethane+isooctane mixture without a polymer, that the correlation length satisfies the asymptotic Ising-like critical behavior, \(\xi=\xi_0t^{-\nu}\), \(t=(T-T_\textup{c})/T_\textup{c}\) with the universal critical exponent \(\nu=0.63\pm 0.01\). This behavior persists for all polymer samples studied and, in disagreement with Ref. \cite{to1998polymer}, we do not find a tendency to the mean-field exponent \(\nu=1/2\). However, the amplitude of the correlation length, \(\xi_0\), depends on the molecular weight and polymer concentration, as shown in  Fig. \ref{fig:Fig3}(b). Moreover, the \(\xi_0\) values obtained for the polystyrene samples follow a universal behavior when the concentration is rescaled as \(cN^2\) (See Section 2 in the Supplemental Material \cite{SupplementalMaterial}).\par
\begin{figure*} 
\centering
\includegraphics[width=4.8in,height=2.1in]{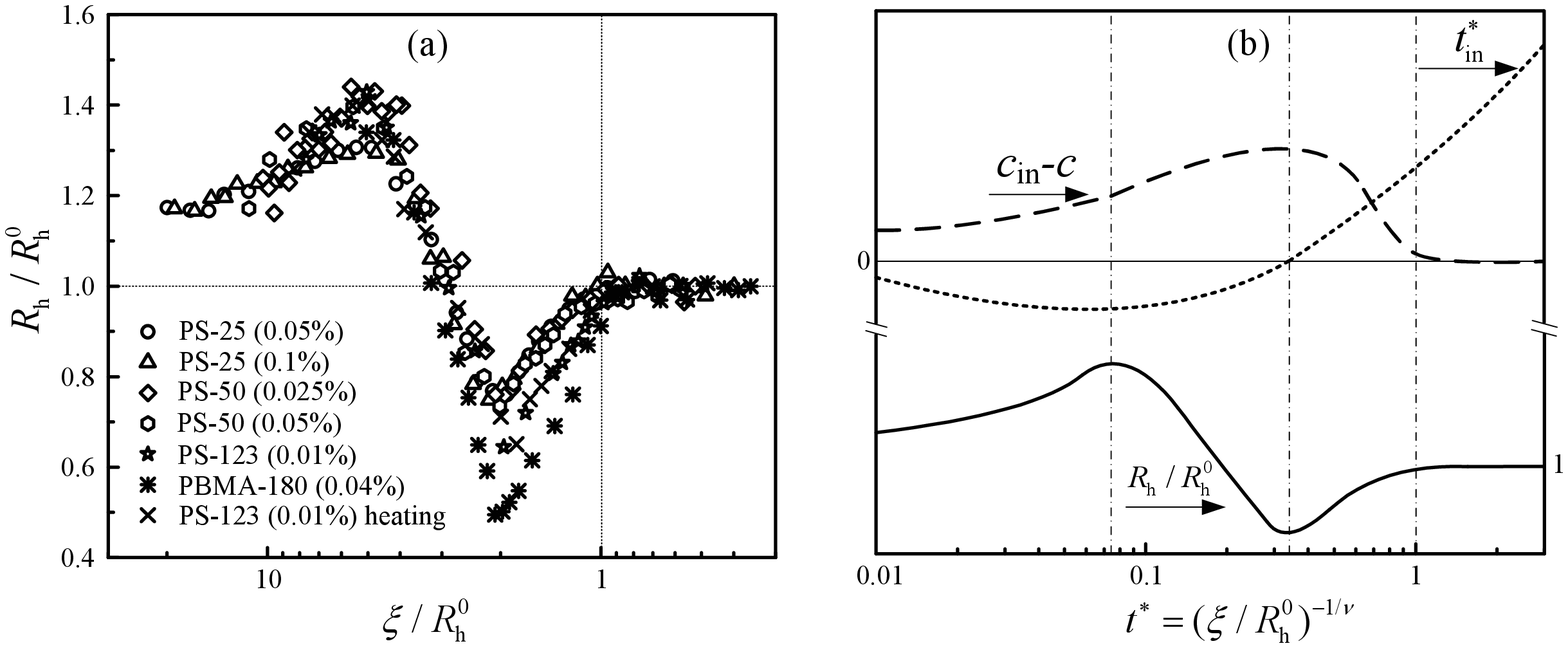}
\caption{\label{fig:Fig4}(a) Scaled relationship between the hydrodynamic radius of polymer coils and the correlation length of the critical fluctuations for different polymer solutions (\(R_\textup{h}^0\) is hydrodynamic radius of a polymer coil for each solution outside the critical region); (b) qualitative interpretation of the collapse-reswelling-expansion-reshrinking transition by the relationship between the scaled hydrodynamic ratio (\(R_\textup{h}/R_\textup{h}^0\), solid curve), the excess concentration inside the coil (\(c_\textup{in}/c\), dashed curve), and the rescaled distance to the critical temperature inside the coil and bulk sample: \(t^*=(T-T_\textup{c})/T_\textup{c}(\xi_\textup{0}/R_\textup{h}^0)^\nu\) and \(t_\textup{in}^*=(T-T_\textup{c})_\textup{in}/T_\textup{c}(\xi_\textup{0}/R_\textup{h}^0)^\nu\) (\(t_\textup{in}^*\), dotted curve).}
\end{figure*}

     From Figs. \ref{fig:Fig1}-\ref{fig:Fig3} we see that all properties strongly depend on the polymer molecular weight and concentration. To investigate the possibility of a better representation of the observed transformation of polymer coils we rescaled the measured hydrodynamic radii and the correlation lengths, reducing them by the ``native'' (far from \(T_\textup{c}\)) hydrodynamic radius, \(R_\textup{h}^0\). The result of such rescaling, shown in Fig. \ref{fig:Fig4}(a), demonstrates a significant degree of universality in the observed phenomenon. The polymer coils begin to collapse when \(\xi/R_\textup{h}^0\approx 1\), reach the minimum size when \(\xi/R_\textup{h}^0\approx 2\) and expand to a maximum size when this ratio is about 5. However, the magnitudes of the collapse and expansion remain dependent on the molecular weight and the nature of polymer.\par
     Two questions arise: (1) What is the connection of the observed first collapse (between \(\xi/R_\textup{h}^0\approx 1\) and \(\xi/R_\textup{h}^0\approx 2\)) and the theory of Brochard and de Gennes? (2) What is the origin of the following expansion and final relaxation to a near-original size at a further approach to the critical temperature? While a coherent theory of this new phenomenon reported in this Letter is yet to be developed, we suggest a plausible interpretation.\par
     Since for a linear-chain molecule the hydrodynamic radius (\(R_\textup{h}\)) is proportional to the radius of gyration (\(R_\textup{g}\)) \cite{teraoka2002polymer}, in our further discussion of the theory we will not make a distinction between these two quantities with respect to how they scale with molecular weight. It is expected that \(R_\textup{g}/R_\textup{h}\) of polystyrene chains is between 1.5 and 1.6 for the nitroethane-isooctane mixture. The fact that the chain collapse begins to be detectable when \(\xi=R_\textup{h}=0.6R_\textup{g}\) is in full agreement with the explanation of Brochard and de Gennes that the collapse begins when the correlation length becomes of the order of the radius of gyration (See details about \(R_\textup{g}/R_\textup{h}\) in Section 3 of the Supplemental Material \citep{SupplementalMaterial})\par
     A simplified explanation of the effect predicted by Brochard and de Gennes is as follows \cite{brochard1980collapse}. The value of the excluded volume   \textit{v} controls the dimensions of polymer coils in a dilute solution. In a good solvent the excluded volume is positive and the coil is swollen. In a poor solvent \(v\) is negative and the coil collapses. However, near the critical point of a good binary solvent the growing correlation length will enhance the preferential adsorption of one of the components, increasing the attraction between the monomers, and reducing the excluded volume. In first approximation, the effective excluded volume \(\tilde{v}\) (proportional to the second virial coefficient in the Flory theory \cite{flory1953principles}) reads:
\begin{equation}
\tilde{v}=v-D^2\chi
\end{equation}
where \(\textit{D}\) is the the strength of the preferential adsorption and \(\chi(\xi, q)=\chi(t, \xi,R_\textup{h})\) is the osmotic susceptibility. If \(\xi/R_\textup{h}^0\ll1\), the susceptibility in the Ornstein-Zernike approximation is \(\chi\propto t^{-2\nu}\propto\xi^2\) \cite{anisimov2005competition}. When \(\xi/R_\textup{h}^0\gg1\), the growth of the susceptibility inside the coil should saturate because of the finite-size effect in the limit \(q\xi\gg1\), \textup{with} \(q\propto R_\textup{h}^{-1}\), as \(\chi(\xi, q)\propto t^{-2\nu}[1+(q\xi)^2]^{-1}\) (not included in the theory of Brochard and de Gennes). Therefore, the first stage of the observed phenomenon (the initial collapse and reswelling) can be qualitatively explained by the theory of Brochard and de Gennes. However, the following significant expansion of the chains, well above the original dimension, contradicts this theory. A possible explanation is elucidated in Fig. \ref{fig:Fig4}(b). We believe that the observed anomaly is caused by micro-phase separation inside the polymer chain. The initial shrinking of polymer chains can indeed be explained in terms of the theory of Brochard and de Gennes: the growing correlation length generates a negative term in the excluded volume. The initial collapse results in the increase of monomer concentration inside the chain and shifting the inside critical temperature into the one-phase region of the bulk sample, so that \((T-T_\textup{c})_\textup{in}<(T-T_\textup{c})\). This effect causes a subsequent transformation of the chain, which is fundamentally different from the reswelling predicted by Brochard and de Gennes. When, in accordance with Eq. (4), the polymer coil starts to collapse, the concentration of monomers in the coil increases, reaching its maximum value at \((T-T_\textup{c})_\textup{in}=0\), meaning that the critical point inside the coil is reached at a positive value of \((T-T_\textup{c})\), at which the bulk phase is still homogeneous. Upon subsequent decrease of the temperature, micro-phase separation will occur inside the coil with one micro-phase enriched with nitroethane (good solvent) and the other one with isooctane (poor solvent). At negative \((T-T_\textup{c})_\textup{in}\) the coil contains a droplet of the poorer solvent (enriched with isooctane), which pushes the coil to expand, possibly due to a critical Casimir force \cite{sumi2009critical}. In the expanded chain, the overall monomer concentration becomes smaller than that in an undisturbed chain. As a result, \((T_\textup{c})_\textup{in}\) returns to \(T_\textup{c}\) of the bulk sample, the micro-phase separation disappears, and the chain reshrinks.\par
According to the theory of Brochard and de Gennes, a polymer coil is ideal far away from the critical point (\(R_\textup{g}\propto N^{1/2}\)), becomes a globule (\(R_\textup{g} \propto N^{1/3}\)) in the most-collapsed state, and becomes ideal again at the critical point. One of the components of our mixed solvent, isooctane, is a typical poor solvent for PS (\(v<0\)) but relatively close to a theta-solvent (\(v=0\)) for PBMA \cite{sander1986solubility}. The other component, nitroethane is likely between a poor-solvent and a theta-solvent for both PS and PBMA. Nevertheless, the PS coils in the mixed solvent demonstrate almost ideal (\(v\approx 0\)) behavior (See Supplemental Material, Section 3 \cite{SupplementalMaterial}). The hydrodynamic radius \(R_\textup{h}\) and the scaling exponent \textit{m} in the power law, \(R_\textup{h}\propto N^m\), for PS coils at different conditions are shown in Table ~\ref{table:radius} and Section 3 of the Supplemental Material \cite{SupplementalMaterial}.\par
\begin{table} 
\centering
\caption{\label{table:radius}Hydrodynamic radius  (\(R_\textup{h}\) nm) and the power-law exponent \textit{m} in molecular-weight scaling of polystyrene coils}
\begin{threeparttable}
\begin{tabular}{lcccc}
   \hline\hline
    & PS-25 & PS-50 & PS-123 & \(\textit{m}\)\\
   \hline
Nitroethane\tnote{1}  & 3.11\(\pm0.30\) & 4.09\(\pm0.40\) & 6.15\(\pm0.60\) & 0.47\(\pm0.01\)\\
Mixed solvent\tnote{1}  & 3.40\(\pm0.35\) & 4.54\(\pm0.40\) & 7.42\(\pm0.68\) & 0.49\(\pm0.01\) \\
Lower phase\tnote{2} & / & / & 6.45\(\pm0.60\) & /\\
Collapsed & 2.55\(\pm0.28\) & 3.40\(\pm0.30\) & 4.78\(\pm0.51\) & 0.39\(\pm0.01\)\\
Expanded & 4.40\(\pm0.20\) & 6.54\(\pm0.20\) & 10.6\(\pm0.51\) & 0.55\(\pm0.01\)\\
   \hline\hline
\end{tabular}
\begin{tablenotes}
        \footnotesize
        \item[1] 50 $^{\circ}$C
        \item[2] in the two-phase region at 25 $^{\circ}$C; the lower phase contains mainly nitroethane
      \end{tablenotes}
    \end{threeparttable}
\end{table}
The fact that the PS coils in the mixed solvent away from the critical point are more ideal than in the pure liquid components is attributed to the effect of co-solvency caused by unfavorable nitroethane-isooctane interactions \cite{magda1988dimensions}. The same effect was also found previously in a mixture of polyethylene oxide with tertiary butanol and water away from the critical point \cite{zheng2018mesoscopic}. At the most-collapsed position, the exponent \textit{m} is close to 1/3, which indicates that the coils tend to become a globule. At the most-expanded position, \textit{m} is greater than 1/2, which means that the coils occupy larger volumes than in the ideal condition. \par
In conclusion, we want to emphasize the unusual nature of the expansion of a polymer coil in the critical region of a mixed solvent well beyond the ideal (\(R_\textup{h}\propto R_\textup{g}\propto N^{1/2}\)) dimension. This expansion is not a well-known swelling of polymer coils in a good solvent (\(R_\textup{g}\propto N^{3/5}\) in the mean-field approximation) \cite{flory1953principles,williams1981polymer}. Instead, the effect is probably caused by liquid-liquid separation inside the coil, which, in turn, is driven by the initial collapse of the coil.\par
     The research of Xiong Zheng was supported by the China Scholarship Council (Grant No. 201606280242), the research of Maogang He was supported by a National Natural Science Fund for Distinguished Young Scholars of China (NSFC No. 51525604). The authors acknowledge logistic and technical support from the Light Scattering Center at the Institute for Physical Science and Technology, the University of Maryland, College Park.
\bibliographystyle{apsrev4-1}
\bibliography{prlNotes.bib}

\end{document}